\documentclass[epj,final]{svjour}
\begin{document}
\title{A quadratic potential in a light cone QCD inspired model}
\author{Hans-Christian Pauli}
\institute{Max-Planck-Institut f\"ur Kernphysik, D-69029 Heidelberg }
\date{20 December 2003}  
\abstract{%
   The general equation from previous work is specialized
   to a quadratic potential $V(r)=-a+\frac12 f r^2$ acting in the space of
   spherically symmetric S wave functions. 
   The fine and hyperfine interaction creates then a position dependent 
   mass $\widetilde m(r)$ in the effective kinetic energy of the associated 
   Schr\"odinger equation.
   The results are compared with the available experimental 
   and theoretical spectral data on the $\pi$ and $\rho$. 
   Solving the eigenvalue problem within the usual oscillator approach 
   induces a certain amount of arbitrariness. 
   Despite of this, the agreement with experimental data is 
   within the experimental error and better than other calculations,
   including Godfrey and Isgur \cite{GodIsg85} 
   and Baldicchi and Prosperi \cite{BalPro02}.
   The short coming can be removed easily in more elaborate work. 
   \PACS{{11.10.Ef}
    \and {12.38.Aw}
    \and {12.38.Lg}
    \and {12.39.-x}
   {}} 
} 
\maketitle
\section{The S-state Hamiltonian}
For spherically symmetric S states the previously derived Hamiltonian
reduces in Fourier approximation to \cite{Pau03a,Pau03b,Pau03c,Pau03d}
\begin{eqnarray}
   \begin{array}{lcl}
   H &=& \phantom{-}
   \frac{\vec p^2}{2m_r} + V + V_{hf}+ V_{K}+ V_{D}  
\,,\\ 
   V_\mathrm{hf} &=& \phantom{-}
   \frac{\vec\sigma_1\vec\sigma_2}{6m_1 m_2} \nabla^2 V   
\,,\\  
   V_\mathrm{K} &=& \phantom{-} 
   \frac{V}{m_1 m_2}  \vec p^2 
\,,\\  
   V_\mathrm{D} &=& -\Big[\frac{V}{16m_1 m_2} \vec p^2 +
   \frac{\nabla^2V}{4m_1 m_2}\Big] \big(\frac{m_1}{m_2}+\frac{m_2}{m_1}\big)  
\,.\end{array}
\label{eq:1}\end{eqnarray}
There are no more interactions than the central potential, 
the hyperfine, the kinetic, and the Darwin interaction, 
but also no less.
For s-states the total spin squared is a good quantum number
$\vec S^2=[(\vec\sigma_1+\vec\sigma_2)/2]^2=S(S+1)$, thus
\begin{eqnarray}
   \vec\sigma_1\vec\sigma_2&=& 2S(S+1) - 3 =
   \left\{\begin{array}{rll}
     +1, & \mbox{ for } S=1, &\mbox{ triplet}, \\
     -3, & \mbox{ for } S=0, &\mbox{ singlet}.\end{array} \right. 
\label{eq:2}\end{eqnarray}
Because it is shorter, $\vec\sigma_1\vec\sigma_2$ is kept 
explicit in the equations as an abbreviation for Eq.(\ref{eq:2}).
With a quadratic potential,
\begin{eqnarray}
   V(r) &=& -a+ \frac{f}{2} r^2 
\,,\end{eqnarray}
the spring constant is $f$, the Hamiltonian (\ref{eq:1}) becomes the 
non-local Schr\"odinger equation
\begin{eqnarray*}
   \begin{array}{rc@{}c@{} l ll ll ll}
   H &=&\big[& \frac{1}{2m_r} &+& \frac{V(r)}{m_1m_2} 
     &-& \frac{V(r)}{16m_1m_2}\big(\frac{m_1}{m_2}+\frac{m_2}{m_1}\big)
     &\big]& \vec p ^2
   \\
     &+&\big[&-a&+&
     \frac{f}{2m_1m_2}\vec\sigma_1\vec\sigma_2
     &-&
     \frac{3f}{4m_1m_2}\big(\frac{m_1}{m_2}+\frac{m_2}{m_1}\big)
     &\big]&+ \frac{f}{2} r^2  
\,,\end{array}
\end{eqnarray*}
since 
\begin{eqnarray}
   \nabla^2 V(r) &=& \frac1r\frac{d^2}{dr^2} rV(r) = 3f
\,.\end{eqnarray}
Shaping notation, the Hamiltonian is written as
\begin{eqnarray}
   H &=& \phantom{-}
   \frac{\vec p^2}{2\widetilde m_r(r)} + \frac{f}{2} r^2 - \widetilde a +
   \widetilde c\, \vec\sigma_1\vec\sigma_2 
\,.\label{eq:5}\end{eqnarray}
The non locality of the Hamiltonian resides in the position
dependent mass
\begin{eqnarray}
   \frac{m_r}{\widetilde m_r(r)} &=&
   1 + \frac{V(r)}{8(m_1+m_2)}
   \Big[16-\frac{m_1}{m_2}-\frac{m_2}{m_1}\Big]
.\end{eqnarray}
To solve this Hamiltonian, one must go on a computer.

The Hamiltonian in Eq.(\ref{eq:5}) 
looks like a conventional instant form Hamiltonian as obtained 
by quantizing the system at equal usual time.
But it must be emphasized that it continues to be 
a genuine front form or light cone Hamiltonian \cite{BroPauPin98},
derived from the latter by a series of exact unitary 
transformations \cite{Pau03a,Pau03c}.

\section{The model Hamiltonian and its parameters}
In this first round, I try to avoid to go on the computer as far
as possible, by the following reason. According to renormalization
theory, the renormalization group invariants (parameters) 
must be determined from experiment. 
This is a strongly non linear problem.
In order to get a first and rough estimate, 
the Hamiltonian is simplified here
until it has a form which is amenable to analytical solution.
Therefore, all in-tractable terms in the above will be replaced 
here by mean values and related to the experimentally accessible 
mean square radius $\langle r^2\rangle$ \cite{PovHue90}. 
In effect, the substitution 
\begin{eqnarray}
   \widetilde m_r(r)\Longrightarrow \widetilde m_r
\,,\label{ieq:8}\end{eqnarray}
is the only true assumption in the present model.
I consider thus the model Hamiltonian, 
\begin{eqnarray}
   H &=& \phantom{-}
   \frac{\vec p^2}{2\widetilde m_r} + \frac{f}{2} r^2 -\widetilde a + 
   \widetilde c\vec\sigma_1\vec\sigma_2 
\,,\label{eq:9}\end{eqnarray}
with the abbreviations 
\begin{eqnarray} 
   \widetilde c &=& 
   \frac{f}{2m_1m_2} 
\,,\\ 
   \widetilde a &=& a + \frac{3\widetilde c }{2}
   \Big(\frac{m_1}{m_2}+\frac{m_2}{m_1}\Big) 
\,,\label{eq:16}\\ 
   \frac{m_r}{\widetilde m_r} &=& 
   1 + \frac{\phantom{8}(f\langle r^2\rangle/2 -a)}{8(m_1+m_2)}
   \Big[16-\frac{m_1}{m_2}-\frac{m_2}{m_1}\Big]
\,.\end{eqnarray}
Its eigenvalues are 
\begin{eqnarray}
   E_n &=& -\widetilde a + \omega\,\xi_0 + \omega\,\eta_n +
   \widetilde c\vec\sigma_1\vec\sigma_2
   \,,\quad 
   \omega = \bigg[\frac{f}{\widetilde m_r}\bigg]^{\frac12} 
,\end{eqnarray}
with $\xi_0=\frac32$ and $\eta_n=2n$.
The invariant mass squares 
\begin{eqnarray}
   M_n^2 &=& \big(m_1+m_2\big)^2  
\nonumber\\ &+&
   2\big(m_1+m_2\big)
   \big(-\widetilde a + \xi_0\omega + \eta_n \omega +
   \widetilde c\vec\sigma_1\vec\sigma_2 \big)
\,,\end{eqnarray}
are then related to experiment.

\begin{table} [t]
   \caption{\label{tab:SingTrip}
   Model parameters in GeV. Note: $[f^*]=1$.
   The heavy flavor masses are determined 
   by the singlet triplet difference.
}\begin{tabular}{||@{\,}c@{\,}||c|c|c|c||c|c||}
 \hline\hline 
 $f^*$  & $m_{d,u}$ & $m_s$ &$m_c$ & $m_b$ &  $a$   &  $100f$   \\ 
 \hline\hline 
  4 & 0.2465 & 0.2648 & 0.2657 & 0.3481 & 0.5412 & 1.7598 
  \\
 \hline\hline
\end{tabular}
\end{table}
For equal masses $m_1=m_2=m$, the model has the 3 parameters 
$m$, $f$ and $a$.
One thus needs 3 empirical data to determine them. I choose:
\begin{eqnarray}
   \begin{array}{rcl}
   M^2_{d \bar u,t1} &=& 4m^2 
   + 4m\big(-\widetilde a+\xi_0\omega + \phantom{3}\widetilde c
   + \eta_1\omega\big)
\,,\\     
   M^2_{d \bar u,t0} &=& 4m^2 
   + 4m\big(-\widetilde a+\xi_0\omega + \phantom{3}\widetilde c\big)
\,,\\     
   M^2_{d \bar u,s0} &=& 4m^2 
   + 4m\big(-\widetilde a+\xi_0\omega - 3\widetilde c\big)
\,.\end{array}
\label{eq:13}\end{eqnarray}
The spectrum is labeled self explanatory by the flavor composition 
$M_n = M_{d\bar u,tn}$ or $M_n = M_{d\bar u,sn}$, 
for singlets or triplets, respectively. 
The triple chosen in Eq.(\ref{eq:13}) exposes a certain asymmetry.
The excited $\rho$ is chosen since its experimental limit 
of error is very much smaller than the one for the corresponding 
$\pi$ state. 
Only its ground state mass  
is known very accurately, \textit{i.e.}
$m_{\pi^+}=139.57018\pm0.00035\mbox{ MeV}$. 
In the present work only the first 4 digits are used.
For equal masses, the above abbreviations become
\begin{eqnarray} 
   \begin{array} {lcl}
   \widetilde c &=& \frac{f}{2m^2} 
   \,, \\  
   \widetilde a &=& a + 3\widetilde c
   \,,\\  
   \frac{m_r}{\widetilde m_r} &=& 
   1 + \frac{7(f\langle r^2\rangle/2 -a)}{8m}
   \,.\end{array}
\end{eqnarray}
The experiment defines 2 certainly positive differences:
\begin{eqnarray}
   \begin{array}{rclcl}
   X^2 &=& \phantom{\frac{\xi_0}{\eta_1}}
   M^2_{d \bar u,t1} - M^2_{d \bar u,t0} &=& 4m
   \phantom{[}\eta_1\omega
\,,\\     
   Y^2 &=& \phantom{\frac{\xi_0}{\eta_1}}
   M^2_{d \bar u,t0} - M^2_{d \bar u,s0} &=& 4m
   \phantom{[}4\widetilde c 
   \,.\end{array}
\label{eq:21}\end{eqnarray}
A third one can be constructed by the observation that
$\frac{\xi_0}{\eta_1}X^2-\frac32 Y^2-M^2_{d \bar u,s0}=4m a - 4m^2$. 
Keeping in mind that 
$\omega ^2= 2f/\widetilde m$,
one can remove trivial kinematic factors and define
3 experimental quantities $B$, $C$ and $D$ by 
\begin{eqnarray} 
   \begin{array}{lcl}
   B^2 &=& \frac14\big(\frac{\xi_0}{\eta_1}X^2 - 
   \frac32 Y^2 - M^2_{d \bar u,s0}\big) = ma-m^2
   \,,\\     
   C^2 &=& 
   \frac{1}{8} Y^2 = \frac{f}{m}
   \,,\\ 
   D^2 &=& \frac{\big(X^2\big)^2}{(4\eta_1)^2Y^2}
   =  
   8m^2 + \frac{7}{2} mf\langle r^2\rangle  - 7ma 
   \,.\end{array}
\label{eq:18}\end{eqnarray}
Substituting $f=mC^2$ and $ma=B^2+m^2$ gives 
\begin{eqnarray*} 
   D^2 &=& \Big[1+\frac72\langle r^2\rangle C^2\Big]m^2 - 7B^2  
\,,\end{eqnarray*}
a quadratic equation with the solution 
\begin{eqnarray}  
   m^2 &=& 
   \frac{D^2+7B^2}{1+\frac{7}{2}C^2\langle r^2\rangle}
.\end{eqnarray}
Having $m$, the $f$ and $a$ are calculated from (\ref{eq:18}). 

\begin{table} [t]
   \caption{\label{tab:model}
   Model parameters in GeV. Note: $[f^*]=1$.
}\begin{tabular}{||@{\,}c@{\,}||c|c|c|c||c|c||}
 \hline\hline 
 $f^*$  & $m_{d,u}$ & $m_s$ & $m_c$  & $m_b$  &  $a$   &  $100f$  \\ 
 \hline\hline 
  2     & 0.4768    & 0.5899 & 1.7646 & 5.1555 & 0.6291 & 3.4035   \\ 
 \hline\hline
\end{tabular}
\end{table}

The position dependent mass changes the relation between the
mean square radius and it experimental value.
Therefore, I introduce a fudge factor $f^*$ according to 
\begin{eqnarray}  
   \langle r^2 \rangle  &=& (f^*)^2 \langle r^2 \rangle_\pi   
\,.\end{eqnarray}
Since all mesons have about the same size \cite{PovHue90}, 
by order of magnitude, this number is kept universal.
The fudge factor is introduced here 
to account, in some global fashion, for the tremendous simplification 
introduced by replacing Eq.(\ref{eq:5}) with (\ref{eq:9}).
Some large scale variations are compiled
in Table~\ref{tab:fudge}. 
The mass spectra including the ground states vary very little with
the fudge factor. Any variations would show up the fastest for 
the high excitations. 
For this reason, the masses for $n=4$ are included in the table.
I do not understand this insensitivity from a mathematical 
or numerical point of view. The major effect of $f^*$ is
the ease by which one can change the quark mass.
A value of $f^*\sim40$ leads to the 20~MeV for the quark mass 
quoted in \cite{BalPro02}. Here $f^*=4$ is chosen.
The other components of the spectrum, both for the $\rho$ 
and the $\pi$, are then obtained for free.
They will be compiled in Table~\ref{tab:ud}, below.
\begin{table}[b]
   \caption{\label{tab:fudge}
   Dependence on the fudge factor.
}\begin{tabular}{||c||@{\ }c@{\ }|@{\ }c@{\ }|@{\ }c@{\ }|@{\ }c@{\ }||c|c||}
\hline\hline
 $f^*$&$m_{d,u}$&$m/\widetilde m $&  $a$  &  $100f$   & 
 4$^1\mathrm{S}_0$ & 4$^3\mathrm{S}_1$\\ 
\hline\hline 
  1.0 &  0.850 &  0.367 &  0.935 &  6.065 & 2.1648 & 2.2929 \\ 
  2.0 &  0.477 &  1.165 &  0.629 &  3.403 & 2.1648 & 2.2929 \\ 
  4.0 &  0.247 &  4.358 &  0.541 &  1.760 & 2.1648 & 2.2929 \\ 
  8.0 &  0.124 & 17.131 &  0.709 &  0.888 & 2.1648 & 2.2929 \\ 
 16.0 &  0.062 & 68.220 &  1.228 &  0.445 & 2.1648 & 2.2929 \\ 
 32.0 &  0.031 & 272.58 &  2.362 &  0.223 & 2.1648 & 2.2929 \\ 
\hline\hline
\end{tabular}
\end{table}

In principle, one could determine the heavier quark masses analytically
from the hyperfine splittings:
\begin{eqnarray} 
   \begin{array}{lclcl}
   \frac1{m_s} &=& \frac{M^2_{u\bar s,t0}-M^2_{u\bar s,s0}}{4f} -\frac1{m_u} 
   \,,\\     
   \frac1{m_c} &=& \frac{M^2_{u\bar c,t0}-M^2_{u\bar c,s0}}{4f} -\frac1{m_u} 
   \,,\\ 
   \frac1{m_b} &=& \frac{M^2_{u\bar b,t0}-M^2_{u\bar b,s0}}{4f} -\frac1{m_u} 
   \,.\end{array}
\end{eqnarray}
The so obtained results are however, not very reasonable, see 
Table~\ref{tab:SingTrip}. The experimental numbers are insufficiently
accurate. Therefore, I determine them 
numerically from the singlets
$M_{u\bar s,s0}$, $M_{u\bar c,s0}$ and $M_{u\bar c,s0}$ and compile 
them in Table~\ref{tab:model}.

\section{Results and Discussion} 
\begin{table}[t]
   \caption{\label{tab:ud}S wave spectra in GeV for light unflavored mesons. 
}\begin{tabular}{||r|l|l||l|l|l||}
 \hline\hline
 n & $^1\mathrm{S}_0$ Singlets& $\pi^+ (u\bar d)$  &n
   & $^3\mathrm{S}_1$ Triplets& $\rho^+(u\bar d)$    \\
   & Experiment$^1$ & Theory    &   & Experiment$^1$& Theory    \\ \hline\hline
 1 & 0.1396(0)      & 0.1396    & 1 & 0.7685(6)     & 0.7685    \\
   &                & 0.1396$^2$&   &               & 0.7711$^2$\\
   &                & 0.150$^3$ &   &               & 0.769$^3$ \\
   &                & 0.497$^4$ &   &               & 0.846$^4$ \\
 2 & 1.300(100)     & 1.2550    & 2 & 1.465(25)     & 1.4650    \\
   &                & 1.2650$^2$&   &               & 1.4650$^2$\\
   &                & 1.300$^3$ &   &               & 0.769$^3$ \\
   &                & 1.326$^4$ &   &               & 1.461$^4$ \\
 3 & 1.795(10)      & 1.7694    & 3 & 1.700(20)$^a$ & 1.9240    \\
   &                & 1.7950$^2$&   &               & 1.9230$^2$\\
   &                & 1.880$^3$ &   &               & 2.000$^3$ \\
   &                & 1.815$^4$ &   &               & 1.916$^4$ \\
 4 & ----           & 2.1648    & 4 & 2.150(17)     & 2.2929    \\
   &                & 2.1620$^2$&   &               & 2.2912$^2$\\
 \hline\hline
 \multicolumn{6}{l}{ %
   $^1$Hagiwara  \textit{etal} \cite{RPP02},\ %
   $^2$Zhou and Pauli \cite{ZhoPau03b}.\ %
}\\
 \multicolumn{6}{l}{ %
   $^3$Godfrey and Isgur \cite{GodIsg85},\ %
   $^4$Baldicchi and Prosperi \cite{BalPro02} (a),\ %
}\\
 \multicolumn{6}{l}{ %
   $^a$Could be a D state \cite{AniAniSar00}.
}\\
\end{tabular}
\end{table}

\textbf{Unflavored light mesons}.
The results for the $\pi$--$\rho$ system 
are compiled in Table~\ref{tab:ud}.
The experimental points are taken from from 
Hagiwara \textit{et al}~\cite{RPP02}. 
It is no surprise that theory and experiment coincide for
the $\pi^+$, the $\rho^+$ and the $\rho^+(1450)$, because
these data have been used to determine the parameters.
The remarkable thing is that one can perform such a fit \emph{at all},
The model reproduces the huge mass of the excited pion within the error limit. 
This solves the long standing puzzle, why a physical system can have a
first excited state with a ten times larger mass.

The remaining three calculated masses of the $\pi$-$\rho$ sector
agree with experiment almost within the error bars.
The model underestimates the second $\pi$-excitation 
by only 26 MeV. The second excitation of the $\rho$
is overestimated by a comparatively large 224 MeV,
but the experiment for the $\rho^+$ ($3^3$S$_1$) needs confirmation. 
The third excitation of the $\rho^+$ ($4^3$S$_1$) is overestimated by 224 MeV.

The table includes also a comparison with other theoretical calculations.
It includes the results from a recent  
oscillator model \cite{ZhoPau03b}. 
Their model is even simpler than the present one: 
it works with a hyperfine splitting, only, 
but suppresses the mechanism of a position dependent mass.
Despite this, the results of \cite{ZhoPau03b} 
coincide practically with the present ones.
I have included also the results from the pioneering work of 
Godfrey and Isgur \cite{GodIsg85} as a prototype of a phenomenological model, 
and from a recent advanced calculation by 
Baldicchi and Prosperi \cite{BalPro02}. 
Neither of these models have much in common with the present one. 
They fail to reproduce the pion, this mystery particle of QCD.

\begin{table}
\caption{\label{table:us} 
   S wave spectra in GeV for strange mesons.}
\begin{tabular}{||l|l@{\,}|l||l|l|l||}
 \hline\hline
 n & Experiment$^1$ & Theory    & n & Experiment$^1$& Theory     \\
 \hline\hline
 \multicolumn{3}{||c@{\,}||}{$^1\mathrm{S}_0$ Singlets $K^+(u\bar s$)}   &
 \multicolumn{3}{  c||}{$^3\mathrm{S}_1$ Triplets $K^{*+}(u\bar s$)}\\
 \hline
 1   & 0.493677(16) & 0.4937    & 1 & 0.89166(26)   & 0.0.8718  \\ 
     &              & 0.6048$^2$&   &               & 0.8917$^2$\\
     &              & 0.47$^3$  &   &               & 0.90$^3$  \\
 2   & 1.460$^a$    & 1.3732    & 2 & 1.629(27)$^b$ & 1.5498    \\ 
     &              & 1.5480$^2$&   &               & 1.6808$^2$\\
     &              & 1.45$^3$  &   &               & 1.58$^3$  \\
 3   & 1.830$^a$    & 1.8782    & 3 & ---           & 2.0109    \\ 
     &              & 2.1040$^2$&   &               & 2.6242$^2$\\
     &              & 2.02$^3$  &   &               & 2.11$^3$  \\
 4   & ---          & 2.2736    & 4 & ---           & 2.3845    \\
 \hline\hline
 \multicolumn{6}{l}{ %
   $^1$Hagiwara  \textit{etal} \cite{RPP02},\ %
   $^2$Zhou and Pauli \cite{ZhoPau03b}.\ %
}\\
 \multicolumn{6}{l}{ %
   $^3$Godfrey and Isgur \cite{GodIsg85},\ %
}\\
 \multicolumn{6}{l}{ %
   $^a$To be confirmed; $^b$$J^P$ not confirmed.
}\\
\end{tabular}
\end{table}

\textbf{Strange mesons}.
The S wave $K^+$ and $K^{*+}$ spectra are given in
Table~\ref{table:us}. The mass of the singlet ground state is
used to determine the mass parameter $m_s$. 
Except the ground states, the experiments carry many ambiguities 
about the quantum number assignment for $K$ and $K^*$ mesons. 
The model prediction for the triplet ground state underestimates
the experimental value by 20 MeV. 
Both the first and the second excited state of $K$
($2^1$S$_0$ and $3^1$S$_0$) are not confirmed. Another unconfirmed
resonance with mass $1.629\pm 0.027$ GeV lying between $2^1$S$_0$
and $3^1$S$_0$ was assigned to be a singlet $K$. Apparently there
is no position for it in the $K$ spectrum if it is an S wave
state. However, according to its mass, it might well be the first
excited state of $K^*$ ($2^1$S$_0$).
Taken the numbers in the table, the discrepancies are
88 and 69 MeV for the singlet and triplet $n=2$ states, respectively.
The second excited state of the $K$ ($2^1$S$_0$) differs by
only 48 MeV, but the datum needs confirmation.
\begin{table}[t]
\caption{\label{table:heavy} Ground state masses in GeV for heavy mesons.}
\begin{tabular}{||l|l@{\,}|l||l|l|l||}
 \hline\hline
 n & Experiment$^1$ & Theory    & n & Experiment$^1$& Theory    \\
 \hline\hline
 \multicolumn{3}{||c||}{$^1\mathrm{S}_0$ Singlet $\bar D^0   (u\bar c)$}  &
 \multicolumn{3}{  c||}{$^3\mathrm{S}_1$ Triplet $\bar D^{*0}(u\bar c)$} \\
 \hline
 1   & 1.8645(5) & 1.9961    & 1 & 2.0067(5)   & 2.0718    \\
     &           & 1.9224$^2$&   &             & 2.0067$^2$\\
     &           & 1.88$^3$  &   &             & 2.04$^3$  \\
 \hline\hline
 \multicolumn{3}{||c||}{$^1\mathrm{S}_0$ Singlet $B^+   (u\bar b)$} &
 \multicolumn{3}{  c||}{$^3\mathrm{S}_1$ Triplet $B^{*+}(u\bar b)$} \\
 \hline
 1   & 5.2790(5) & 5.2790    & 1 & 5.3250(6)   & 5.3085    \\
     &           & 5.2965$^2$&   &             & 5.3250$^2$\\
     &           & 5.31$^3$  &   &             & 5.37$^3$  \\
 \hline\hline
 \multicolumn{3}{||c||}{$^1\mathrm{S}_0$ Singlet $D^-_s   (s\bar c)$}  &
 \multicolumn{3}{  c||}{$^3\mathrm{S}_1$ Triplet $D^{*-}_s(s\bar c)$} \\
 \hline
 1   & 1.9685(6) & 1.9961    & 1 & 2.1124(7)   & 2.0718    \\
     &           & 2.0201$^2$&   &             & 2.0655$^2$\\
     &           & 1.98$^3$  &   &             & 2.13$^3$  \\
 \hline\hline
 \multicolumn{3}{||c||}{$^1\mathrm{S}_0$ Singlet $B^0_s   (s\bar b)$}  &
 \multicolumn{3}{  c||}{$^3\mathrm{S}_1$ Triplet $B^{*0}_s(s\bar b)$} \\
 \hline
 1   & 5.3696(24) & 5.3976    & 1 & 5.4166(35)  & 5.4214    \\
     &            & 5.3739$^2$&   &             & 5.3885$^2$ \\
     &            & 5.35$^3$  &   &             & 5.45$^3$  \\
 \hline\hline
 \multicolumn{3}{||c||}{$^1\mathrm{S}_0$ Singlet $B^+_c   (c\bar b)$} &
 \multicolumn{3}{  c||}{$^3\mathrm{S}_1$ Triplet $B^{*+}_c(c\bar b)$} \\
 \hline
 1   & 6.4(4) & 6.5077   & 1 & --- & 6.5157     \\
     &        & ---      &   &     & 6.3458$^2$ \\
     &        & 6.27$^3$ &   &     & 6.34$^3$   \\
 \hline\hline
 \multicolumn{6}{l}{ %
   $^1$Hagiwara  \textit{etal} \cite{RPP02},\ %
   $^2$Zhou and Pauli \cite{ZhoPau03b}.\ %
}\\
 \multicolumn{6}{l}{ %
   $^3$Godfrey and Isgur \cite{GodIsg85},\ %
}\\
 \multicolumn{6}{l}{ %
   $^a$To be confirmed; $^b$$J^P$ not confirmed.
}\\
\end{tabular}
\end{table}

\textbf{Heavy mesons}.
The S wave $u\bar c$, $u\bar b$, $s\bar c$, $s\bar b$ and $c\bar
b$ meson spectra are given in Table~\ref{table:heavy}. No
excitations were observed for these mesons. 

The $u\bar c$ singlet is used to determine the mass parameter $m_c$.
Its ground state mass ($\bar D^{0}$) therefore coincides with experiment. 
The model overestimates the mass of the triplet $\bar D^{*0}$ ($1^1$S$_0$)
by about 50 MeV.~---
The $u\bar b$ singlet is used to determine the mass parameter $m_b$.
Its ground state mass ($\bar B^{+}$) therefore coincides with experiment. 
The model overestimates the mass of the triplet $\bar B^{*+}$ ($1^1$S$_0$)
by 16 MeV only. 

No data in the $s\bar c$ mesons are used to determine model parameters. 
Model and experiment differ by 27 and 40 MeV 
for singlet and triplet, respectively.

Model and experiment differ by 27 and 4 MeV 
for singlet and  triplet, respectively. 

Model and experiment agree for the singlet.
The triplet data are unknown. 

The model prediction for the complete spectrum are compiled
in Table~\ref{tab:specHeavy}, for easy reference.

The flavor diagonal mesons like $s\bar s$, $s\bar c$ or $b\bar b$ 
may not be calculated in the model, see \cite{BroPauPin98}.

\section{Conclusions}

The agreement between the present simple model with an 
oscillator potential and the experiment is generally good. 
These are good news, since harmonic interactions are easy to work with
in many body problems. The present approach will be useful
for considering baryons and nuclei.

With the 4 mass parameters of the up/down, strange, charm and
bottom quarks, the model has only 2 additional 2 parameters 
for the harmonic oscillator potential. In principle, the fudge
factor must be counted as parameter, but as seen above, the choice
of the up/down and the fudge factor is strongly coupled.

The 6 canonical parameters of the model generate a reasonably good
agreement with the 21 data points available.

Note that renormalized gauge field theory has also 4+1+1
parameters:
The 4 flavor quark masses, the strong coupling constant $\alpha_s$,
and the renormalization scale $\lambda$.
Of course, they can be mapped into each other \cite{Pau03a,Pau03b,Pau03c}.

Once one has determined the parameters in such a first guess, 
one should relax the model assumption, Eq.(\ref{ieq:8}),
and work with the full non local model, with a position
dependent mass. For this one has to go back to the computer
and perform the necessary fine tunings of the parameters. 

\begin{table}[t]
\caption{\label{tab:specHeavy}
   The predicted S spectrum in GeV for heavy mesons.
}\begin{tabular}{||l|l@{\hspace{9ex}}||l|l@{\hspace{9ex}}||}
 \hline\hline
 n$^1\mathrm{S}_0$  & $\bar D^0   \qquad(u\bar c)$ &
 n$^3\mathrm{S}_1$  & $\bar D^{*0}\qquad(u\bar c)$ \\
 \hline
    1 & 1.8645 &  1  & 1.9594 \\
    2 & 2.5013 &  2  & 2.5728 \\
    3 & 3.0061 &  3  & 3.0658 \\
    4 & 3.4375 &  4  & 3.4899 \\
 \hline\hline
 n$^1\mathrm{S}_0$  & $B^+   \qquad(u\bar b)$ &
 n$^3\mathrm{S}_1$  & $B^{*+}\qquad(u\bar b)$ \\
 \hline
    1 & 5.2790 &  1  & 5.3085 \\
    2 & 5.8473 &  2  & 5.8739 \\
    3 & 6.3651 &  3  & 6.3896 \\
    4 & 6.8438 &  4  & 6.8666 \\
 \hline\hline
 n$^1\mathrm{S}_0$  & $D^-_s   \qquad(s\bar c)$ &
 n$^3\mathrm{S}_1$  & $D^{*-}_s\qquad(s\bar c)$ \\
 \hline
    1 & 1.9961 &  1  & 2.0718 \\
    2 & 2.5837 &  2  & 2.6426 \\
    3 & 3.0605 &  3  & 3.1104 \\
    4 & 3.4724 &  4  & 3.5165 \\
 \hline\hline
 n$^1\mathrm{S}_0$  & $B^0_s   \ \qquad(s\bar b)$ &
 n$^3\mathrm{S}_1$  & $B^{*0}_s\ \qquad(s\bar b)$ \\
 \hline
    1 & 5.3976 &  1  & 5.4214 \\
    2 & 5.9163 &  2  & 5.9380 \\
    3 & 6.3930 &  3  & 6.4131 \\
    4 & 6.8365 &  4  & 6.8553 \\
 \hline\hline
 n$^1\mathrm{S}_0$  & $B^+_c   \qquad(c\bar b)$ &
 n$^3\mathrm{S}_1$  & $B^{*+}_c\qquad(c\bar b)$ \\
 \hline
    1 & 6.5077 &  1  & 6.5157 \\
    2 & 6.8447 &  2  & 6.8523 \\
    3 & 7.1659 &  3  & 7.1731 \\
    4 & 7.4733 &  4  & 7.4802 \\
 \hline\hline
\end{tabular}
\end{table}
\end{document}